\documentstyle[prl,aps,epsfig]{revtex}         

\begin{document}
\catcode`\ä = \active \catcode`\ö = \active \catcode`\ü = \active
\catcode`\Ä = \active \catcode`\Ö = \active \catcode`\Ü = \active
\catcode`\ß = \active \catcode`\é = \active \catcode`\è = \active
\catcode`\ë = \active \catcode`\ô = \active \catcode`\ê = \active
\catcode`\ø = \active \catcode`\ò = \active \catcode`\í = \active
\defä{\"a} \defö{\"o} \defü{\"u} \defÄ{\"A} \defÖ{\"O} \defÜ{\"U} \defß{\ss} \defé{\'{e}}
\defè{\`{e}} \defë{\"{e}} \defô{\^{o}} \defê{\^{e}} \defø{\o} \defò{\`{o}} \defí{\'{i}}
\draft               
\newcommand{\li}{$^6$Li}
\newcommand{\na}{$^{23}$Na}
\twocolumn [\hsize\textwidth\columnwidth\hsize\csname
@twocolumnfalse\endcsname 

\title{Two-species mixture of quantum degenerate Bose and Fermi gases}
\vspace{-5mm}
\author{Z. Hadzibabic, C. A. Stan, K. Dieckmann, S. Gupta, M. W. Zwierlein, A. G\"{o}rlitz, and W. Ketterle}
\address{Department of Physics, MIT-Harvard Center for Ultracold
Atoms, and Research Laboratory
of Electronics, \\
MIT, Cambridge, MA 02139}
\date{\today}
\maketitle

\begin{abstract}
We have produced a macroscopic quantum system in which a {\li}
Fermi sea coexists with a large and stable {\na} Bose-Einstein
condensate. This was accomplished using inter-species sympathetic
cooling of fermionic {\li} in a thermal bath of bosonic {\na}.
\end{abstract}
\vskip1pc ]

\narrowtext

Experimental methods of laser and evaporative cooling, used in the
production of atomic Bose-Einstein condensates (BEC) \cite
{ingu99}, have recently been extended to realize quantum
degeneracy in trapped Fermi gases \cite
{dema99,trus01,schr01,gran01}. What makes gaseous fermionic
systems particularly appealing to investigate is the relative ease
with which their properties can be varied. This allows the
exploration of a vast range of experimental regimes, from
non-interacting to strongly correlated. In the first case, purely
quantum statistical effects can be studied, such as the
implications of Pauli exclusion on scattering properties of the
system. In the other extreme, exciting new regimes of BCS-like
fermionic superfluidity may be within reach \cite
{stoo96,houb97super,bara98,holl01}. An additional area of interest
is the production of a dilute quantum degenerate mixture of Bose
and Fermi gases, akin to the strongly interacting
$^{4}$He$\,$-$^{3}$He liquid. This would extend the list of
possible experimental studies even further, to include effects
such as interaction-driven phase separation \cite{molm98zero} or
BEC induced interactions between fermions.

In this Letter, we report the production of a new macroscopic
quantum system, in which a degenerate {\li} Fermi gas coexists
with a large and stable {\na} BEC. We have achieved high numbers
of both fermions ($>10^{5}$) and bosons ($>10^{6}$), and {\li}
quantum degeneracy corresponding to one half of the Fermi
temperature ($T_{F}$). Low rates for both intra- and inter-species
inelastic collisions result in a lifetime longer than $10\,$s.
Hence, in addition to being the starting point for studies of the
degenerate Fermi gas, this system shows great promise for studies
of degenerate Bose-Fermi mixtures, including collisions between
the two species, and limitations to the cooling process\cite
{timm98super}. Further, we point out that {\li} is regarded as a
particularly promising candidate for the BCS transition \cite
{stoo96,houb97super}.

Our experimental approach is based on sympathetic cooling of
fermions in a large bosonic ``refrigerator''. In contrast to the
bosonic case, two-body elastic collisions are absent in a
single-component Fermi gas at ultra-low temperatures due to the
Pauli exclusion principle. This lack of thermalization precludes
direct implementation of forced evaporative cooling. Therefore,
cooling of fermions into the quantum degenerate regime must rely
on collisions between distinguishable atoms. In two experiments
which produced degenerate Fermi gases, mixtures of two fermionic
spin states were simultaneously evaporated and mutually cooled
\cite {dema99,gran01}. Two groups have also demonstrated
sympathetic cooling of {\li} by the $^{7}$Li bosonic isotope, thus
also producing the first quantum degenerate Bose-Fermi mixtures.
However, this system has a limitation that in the upper hyperfine
state, the $^{7}$Li condensate is unstable \cite {trus01}, while
scattering properties in the lower hyperfine state make
sympathetic cooling inefficient, and limit the size of both {\li}
and $^{7}$Li samples \cite {schr01}. We have overcome both of
these limitations by using a large {\na} cloud, instead of
$^{7}$Li, for sympathetic cooling of {\li}. Our work provides the
natural progression in the search for an ideal Bose-Fermi system,
where a ``good'' Bose-Einstein condensate is chosen, and then
combined with a favorable fermionic species. Similar two-species
experiments are currently being pursued by two other groups \cite
{gold01,modu01}. Given the vast variety of collisional properties
among alkali gases, and a limited choice of favorable Bose-Fermi
combinations, the properties of the {\li} - {\na} mixture are
truly fortuitous. Both the presence of sufficient ``good''
(elastic) collisions needed for inter-species thermalization, and
the slow rate of ``bad'' (inelastic) collisions could not be taken
for granted before our studies. It is also worth noting that in
our experiment, a mixture of two different atomic species has been
simultaneously brought into quantum degeneracy for the first time.

For this experiment, we have upgraded our {\na} BEC apparatus
\cite {mewe96bec} to allow for both lithium and sodium operation,
while making minimal modifications to the original setup. The
additional laser light needed for optical cooling of {\li} was
generated by a low power, diode laser system \cite {schu98}. The
lithium (671$\,$nm) and the sodium ($589\,$nm) laser beams were
overlapped using dichroic beam-splitters, and the spatial
arrangement of the laser and atomic beams used to trap, cool, and
detect lithium was identical to the original sodium setup.
Specifically, a two-species magneto-optical trap (MOT) was loaded
from a single two-species atomic beam, slowed in the same Zeeman
slower previously used in sodium-only experiments. The fact that
the maximum spontaneous light force deceleration is twice as large
for {\li} as for {\na} allowed us to slow lithium atoms without
compromising the slowing efficiency for sodium.

\begin{figure}[htbf]
\begin{center}
\vskip-4mm \epsfxsize=80mm {\epsfbox{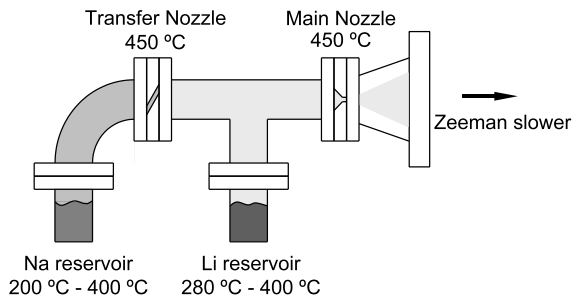}} \vskip-7mm
\end{center}
\caption{Double Species Oven. {\li} and {\na} vapors were produced
in separate chambers to allow for independent control of the atom
fluxes. The two species were mixed in the lithium chamber. The
transfer nozzle has a conductance 40 times lower than the main
nozzle, and limits the undesirable diffusion of lithium into the
sodium chamber.} \label{fig:DoubleOven}
\end{figure}

To implement our experimental strategy, we have developed a double
species oven in which the vapors of {\li} and {\na} were mixed,
and a single atomic beam containing both species was produced
(Fig.$\,$\ref{fig:DoubleOven}). The main difficulty in designing
such an oven is that at the same temperature, the vapor pressure
of lithium is three orders of magnitude lower than that of sodium.
To achieve comparable atomic fluxes of both species, the alkali
vapors must be produced in separate chambers, and then delivered
to a mixing chamber, at controllable rates. In our design, the
lithium chamber was also used for mixing. To operate the oven in
either single or double species mode, we tuned the atomic fluxes
independently by changing the temperatures of the alkali
reservoirs. The maximum atom fluxes into the solid angle subtended
by the  MOT region were $3 \times 10^{11}\,$s$^{-1}$ for {\li},
and $2 \times 10^{12}\,$s$^{-1}$ for {\na}.

Under typical operating conditions, $5\,$s of loading resulted in
single species MOTs with $2 \times 10^{7}$ lithium atoms or $6
\times 10^{9}$ sodium atoms. When both MOTs were operated
simultaneously, inter-species light assisted collisions reduced
the number of lithium atoms by a factor of about four, while the
sodium atom number was not noticeably affected. The number of
{\li} atoms in the MOT was maximized when the trapping and the
repumping light frequencies were tuned 25$\,$MHz below the
corresponding resonances. The resulting temperature of the lithium
atoms was $\sim 700\,\mu$K.

\begin{figure}[htbf]
\begin{center}
\vskip 3mm \epsfxsize=85mm {\epsfbox{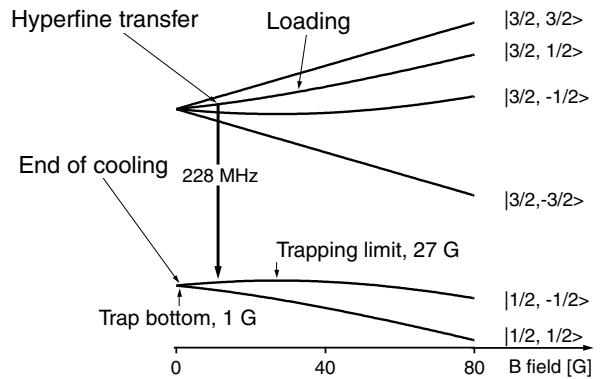}} \vskip -3mm
\end{center}
\caption{{\li} in the magnetic trap. Ground state energy levels:
The six hyperfine states are labelled in the low magnetic field,
$|F,m_{F}\rangle$ basis. The $|1/2,-1/2\rangle$ state becomes
strong field seeking for fields above $27\,$G, limiting the trap
depth to $330\,\mu$K. Cooling path: The atoms are loaded into the
magnetic trap in the upper hyperfine state, at a temperature of
$700\,\mu$K. After the initial cooling stage to $50\,\mu$K, the
atoms are transferred into the $|1/2,-1/2\rangle$ state, and
further cooled to a final temperature of $330\,$nK.}
\label{fig:LiBreit-Rabi}
\end{figure}

Since the {\na} BEC is produced in the
$|F,m_{F}\rangle=|1,-1\rangle$ lower hyperfine ground state, to
avoid inelastic spin-exchange collisions, it is preferred to
magnetically trap {\li} in the corresponding $|1/2,-1/2\rangle$
state. (Here, $F$ is the total angular momentum, and $m_{F}$ is
its projection along the quantization axis.) However, the maximum
magnetic trap depth in the $|1/2,-1/2\rangle$ state is only
$330\,\mu$K (see Fig.$\,$\ref{fig:LiBreit-Rabi}), considerably
lower than our MOT temperature. Further, due to inefficiency of
sub-Doppler cooling mechanisms, it is not possible to optically
cool lithium to temperatures which would make magnetic trapping in
this state efficient\cite {schu98}. Therefore, to avoid drastic
losses due to the limited trap depth, lithium atoms were optically
pumped and then magnetically trapped in the $F=3/2$ manifold.
Before loading the magnetic trap, $4\,$ms were allowed for
sub-Doppler polarization gradient cooling of sodium, during which
the lithium cloud was in free expansion. This reduced the transfer
efficiency of lithium atoms into the trap by a factor of 2,
limiting it to about 12$\,\%$. We have thus magnetically trapped
$\sim 6 \times 10^{5}$ {\li} atoms in the upper hyperfine state,
and $\sim 2 \times 10^{9}$ {\na} atoms in the lower one. At low
energies, our cloverleaf magnetic trap is harmonic and axially
symmetric. In the lower hyperfine states, the trapping frequencies
for lithium (sodium) are $\omega_{z} = 2\pi \,\times$ 26 (16) Hz
axially, and $\omega_{r} = 2\pi \,\times$ 354 (221) Hz radially.

Once the atoms were loaded into the magnetic trap, we started the
forced evaporative cooling of sodium \cite {ingu99}. A varying
microwave field near 1.77$\,$GHz was used to gradually lower the
trap depth for {\na} by selectively transferring the most
energetic atoms into the untrapped $|2,-2\rangle$ state. This
microwave field does not affect the {\li} atoms, which are
therefore not evaporated. Cooling of the lithium sample is instead
achieved through thermal contact with sodium. We have observed
efficient sympathetic cooling of {\li} in the {\it upper}
hyperfine state by {\na} in the {\it lower} one, and have
successfully cooled this Bose-Fermi mixture into simultaneous
quantum degeneracy. This observation indicates a surprisingly
favorable ratio between the good and the bad inter-species
collisions in this mixture. The losses due to inelastic
spin-exchange collisions took place only on a time scale of
several seconds, comparable to the total evaporation time of
$15\,$s.

In order to produce a collisionally stable Bose-Fermi mixture, it
is necessary to transfer the lithium atoms into the lower
hyperfine state. To minimize the initial losses due to
spin-exchange collisions, this transfer should take place as early
in the cooling process as possible. On the other hand, before
lithium atoms can be efficiently trapped in the lower hyperfine
state, they must be cooled significantly below $330\,\mu$K.
Therefore, we implemented sympathetic cooling in two stages (see
Fig.$\,$\ref{fig:LiBreit-Rabi}). We have optimized the initial
evaporation stage to reach a temperature of $\sim 50\,\mu$K in
$5\,$s, while losing less than half of the lithium atoms, and
maintaining the conditions for efficient sodium evaporation. At
this point, we found that a substantial fraction of lithium atoms
was in the $|3/2,1/2\rangle$ state. They could thus be transferred
to the $|1/2,-1/2\rangle$ state on a single-photon RF transition
at 228$\,$MHz, which is, to first order, independent of the
magnetic field. This simplification over a similar hyperfine
transfer previously employed in \cite {schr01} was not expected.
After the RF pulse was applied, the remaining $F=3/2$ atoms were
optically pumped into untrapped hyperfine states, and expelled
from the trap. If this ``clean-up'' light pulse was omitted,
spin-exchange collisions between lithium atoms in different
hyperfine states led to a rapid loss of atoms from the trap. The
overall efficiency of our hyperfine transfer was $\sim 50\,\%$.
The evaporation was then resumed for another $10\,$s. We have
observed efficient sympathetic cooling of the $|1/2,-1/2\rangle$
atoms, and have cooled both gases into quantum degeneracy without
observable losses in the lithium atom number.

\begin{figure}[htbf]
\begin{center}
\epsfxsize=\columnwidth {\epsfbox{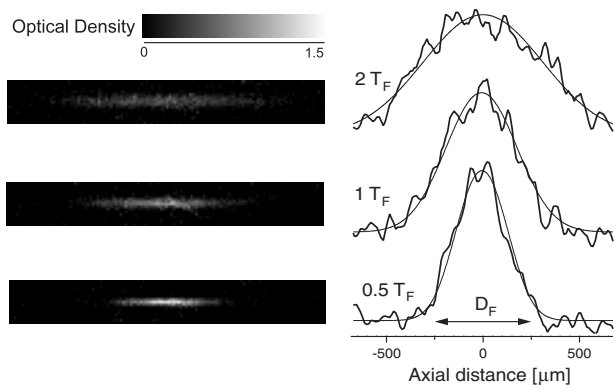}}
\end{center}
\caption{Onset of Fermi degeneracy. Three pairs of images (top to
bottom) correspond to $T/T_{F} =$ 2, 1, and 0.5. (a) Column
densities of the {\li} cloud were recorded by absorption imaging.
(b) Axial line density profiles and the Fermi-Dirac fits to the
data are plotted. The arrow indicates the size of the Fermi
diameter, $D_{F}$.} \label{fig:Waterfall}
\end{figure}

Fig.$\,$\ref{fig:Waterfall}(a) displays the effect of sympathetic
cooling on the {\li} cloud. Absorption images of the trapped {\li}
gas were taken after the {\na} evaporation was terminated at
different trap depths, and the sample was allowed to fully
thermalize. Cooling (from top to bottom) is seen in the shrinking
of the density distribution, and an increase in the peak optical
density. In contrast to standard evaporative cooling, and the
mutual cooling between two Fermi species, the total number of
atoms remains constant.

Quantitative analysis of the {\li} clouds is depicted in
Fig.$\,$\ref{fig:Waterfall}(b). We have performed two-dimensional
fits to the recorded column densities using both a simple Gaussian
model, and an exact semi-classical Fermi-Dirac distribution \cite
{butt97}. While the former is a valid description of the gas only
in the classical regime, the latter approach is valid at all
temperatures. Indeed, at higher temperatures, the two fits
performed equally well, and yielded the same temperature. However,
at a temperature of about 400$\,$nK, the classical fits started to
fail. This was indicated by the relative growth of the reduced
$\chi^{2}$ values, by up to $20\,\%$ above the corresponding
values for the Fermi-Dirac fits. For the coldest samples, Gaussian
fits also overestimate the temperature by $\sim 15\,\%$. From the
fitted number of atoms in the system ($\sim 1.5 \times 10^{5}$),
we found that the noticeable inadequacy of the classical fits
occurs at $\sim 0.6\,T_{F}$, and is a clear signature of the Fermi
degeneracy. Fig.$\,$\ref{fig:Waterfall}(b) shows projected line
densities along the axial direction of the {\li} cloud, and the
Fermi-Dirac fits to the data. The arrow indicates the size of the
Fermi diameter, $D_{F} = 2 \sqrt{2 k_{B} T_{F}/(m
\omega_{z}^{2})}$, for the fitted atom number. The spatial extent
of the coldest cloud is already comparable to this minimum size
the system would assume at zero temperature. The coldest {\li}
samples are produced in coexistence with almost pure {\na}
condensates with $\sim 2 \times 10^{6}$ atoms. The lifetime of
this degenerate mixture is limited to about 10$\,$s by the
three-body decay of the BEC, while the lithium cloud has a
lifetime longer than 100$\,$s.

\begin{figure}[htbf]
\begin{center}
\epsfxsize=80mm {\epsfbox{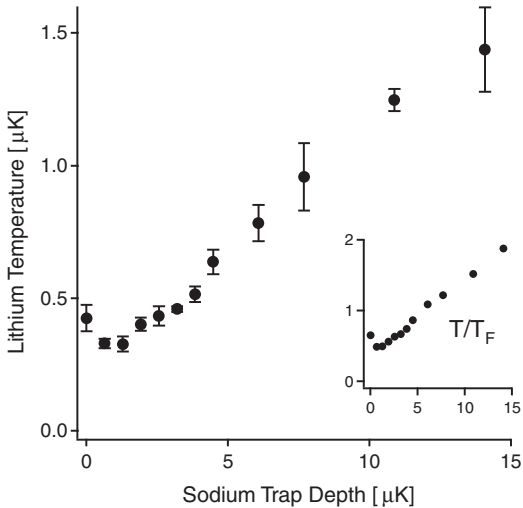}}
\end{center}
\caption{Temperatures of the {\li} cloud as a function of the
{\na} trap depth. Each data point is an average of three
measurements. The error bars indicate the shot-to-shot
fluctuations, while the uncertainties of the fits are comparable
or smaller. In the case when sodium was completely evaporated from
the trap ({\it c.f.} zero trap depth data point), the very last
stage of sympathetic cooling became inefficient due to the
vanishing heat capacity of the bosonic reservoir. The insert shows
the same temperature data scaled to the Fermi temperature.}
\label{fig:RFscan}
\end{figure}

In Fig.$\,$\ref{fig:RFscan}, the temperature of the {\li} cloud is
plotted as a function of the final {\na} trap depth. The highest
observed Fermi degeneracy is given by $T \approx 330\,$nK $\approx
0.5\,T_{F}$. We have also compared the {\li} temperatures with the
temperatures of the {\na} cloud, extracted from the thermal wings
of the bosonic density distribution. In hotter samples, the two
agreed to within $10\,\%$. This indicates that the inter-species
thermalization in our experiment took place on time scales shorter
than a second. Further quantitative study of the thermalization
rate and the inter-species elastic cross-section was not
attempted. For the coldest samples, we observe that the equality
of the {\li} and the {\na} temperature does not hold any more. The
lowest measurable temperature of sodium was $T \approx 170\,$nK,
about half the corresponding lithium value, and we observed even
purer condensates for which the temperature could not be
accurately determined. We have verified that extending the
thermalization time at the end of evaporation up to 10$\,$s did
not lower the lithium temperature any further. Spatial separation
of the two clouds due to different gravitational sags can also be
readily ruled out. Therefore, if a common temperature for the two
species is assumed, this would imply a {\li} Fermi degeneracy
corresponding to $0.25\,T_{F}$, comparable to the values
previously obtained under this assumption \cite {trus01,schr01}.
However, while it is worth noting that in the deeply degenerate
regime the Fermi-Dirac distribution is almost insensitive to
temperature variations, we regard the more conservative
temperatures obtained from the Fermi-Dirac fits as reliable. The
reasons for the observed temperature discrepancy are worthy of
further investigation.

In conclusion, we have produced a system in which a {\li} Fermi
sea coexists with a {\na} BEC. This provides us with a starting
point for studies of the degenerate Bose-Fermi mixtures. In
particular, the observed temperature difference between the two
spatially overlapped species might provide further insight into
the limits of sympathetic cooling. Further, by loading the
degenerate Fermi cloud into an optical trap, effects of
magnetically tunable interactions between lithium atoms in
different spin states can be studied \cite {houb98fesh}. A
particularly appealing prospect is the observation of the BCS
transition to a fermionic superfluid state, for which {\li} is a
very promising candidate.

We would like to thank Florian Schreck for useful discussions and
Christian Schunck for experimental assistance. This research was
supported by NSF, ONR, ARO, NASA, and the David and Lucile Packard
Foundation. M. W. Z. acknowledges the support of the
Studienstiftung des deutschen Volkes and the Ecole Normale
Sup\'{e}rieure, Paris.

\vspace{-0.7cm}
\bibliographystyle{prsty}
\bibliography{References,ReferencesNew}

\begin{thebibliography}{10}

\bibitem{ingu99}
M. Inguscio, S. Stringari, and C.~E. Wieman, {\em Bose-Einstein condensation in
  atomic gases, Proceedings of the International School of Physics Enrico
  Fermi, Course CXL} (IOS Press, Amsterdam, 1999), (editors).

\bibitem{dema99}
B. DeMarco and D.~S. Jin, Science {\bf 285},  1703  (1999).

\bibitem{trus01}
A.~G. Truscott, K.~E. Strecker, W.~I. McAlexander, G.~B. Partridge, and R.~G.
  Hulet, Science {\bf 291},  2570  (2001).

\bibitem{schr01}
F. Schreck, L. Khaykovich, K.~L. Corwin, G. Ferrari, T. Bourdel, J. Cubizolles,
  and C. Salomon, Phys. Rev. Lett. {\bf 87},  080403  (2001).

\bibitem{gran01}
S.~R. Granade, M. Gehm, K.~M. O'Hara, and J.~E. Thomas, cond-mat/0111344  .

\bibitem{stoo96}
H.~T.~C. Stoof, M. Houbiers, C.~A. Sackett, and R.~G. Hulet, Phys. Rev. Lett.
  {\bf 76},  10  (1996).

\bibitem{houb97super}
M. Houbiers, R. Ferwenda, H.~T.~C. Stoof, W. McAlexander, C.~A. Sackett, and
  R.~G. Hulet, Phys. Rev. A {\bf 56},  4864  (1997).

\bibitem{bara98}
M.~A. Baranov and D.~S. Petrov, Phys. Rev. A {\bf 58},  R801  (1998).

\bibitem{holl01}
M. Holland, S.~J. J. M.~F. Kokkelmans, M.~L. Chiofalo, and R. Walser, Phys.
  Rev. Lett. {\bf 87},  120406  (2001).

\bibitem{molm98zero}
K. Mølmer, Phys. Rev. Lett. {\bf 80},  1804  (1998).

\bibitem{timm98super}
E. Timmermans and R. Côté, Phys. Rev. Lett. {\bf 80},  3419  (1998).

\bibitem{gold01}
J. Goldwin, S.~B. Papp, B. DeMarco, and D.~S. Jin, submitted to Phys. Rev. A  .

\bibitem{modu01}
G. Modugno, G. Ferrari, G. Roati, R.~J. Brecha, A. Simoni, and M. Inguscio,
  Science {\bf 294},  1320  (2001).

\bibitem{mewe96bec}
M.~O. Mewes, M.~R. Andrews, N.~J. van Druten, D.~M. Kurn, D.~S. Durfee, and W.
  Ketterle, Phys. Rev. Lett. {\bf 77},  416  (1996).

\bibitem{schu98}
U. Schünemann, H. Engler, M. Zielonowski, M. Weidemüller, and R. Grimm, Optics
  Comm. {\bf 158},  263  (1998).

\bibitem{butt97}
D.~A. Butts and D.~S. Rokhsar, Phys. Rev. A {\bf 55},  4346  (1997).

\bibitem{houb98fesh}
M. Houbiers, H.~T.~C. Stoof, W. McAlexander, and R.~G. Hulet, Phys. Rev. A {\bf
  57},  R1497  (1998).

\end{thebibliography}

\end{document}